\documentclass[10pt, conference]{IEEEtran}
\IEEEoverridecommandlockouts

\usepackage{cite}
\usepackage{amsmath,amssymb,amsfonts}
\usepackage{algorithmic}
\usepackage{graphicx}
\usepackage{textcomp}
\usepackage{xcolor}
\usepackage{booktabs} 
\usepackage{tablefootnote}
\usepackage{multirow}
\usepackage{subcaption}
\usepackage{enumitem}
\setlist[itemize]{noitemsep, nolistsep}
\def\BibTeX{{\rm B\kern-.05em{\sc i\kern-.025em b}\kern-.08em
    T\kern-.1667em\lower.7ex\hbox{E}\kern-.125emX}}
\begin{document}


\setlength{\textfloatsep}{5pt} 
\setlength{\floatsep}{5pt}     
\setlength{\belowcaptionskip}{0pt} 
\setlength{\abovedisplayskip}{0pt}

\title{Hybrid Systolic Array Accelerator with Optimized Dataflow for Edge Large Language Model Inference   \\
}


\author{\IEEEauthorblockN{Chun-Ting Chen, HanGyeol Mun, Jian Meng, Mohamed S. Abdelfattah, and Jae-sun Seo\thanks{Accepted as a conference paper in  International Symposium on Low Power Electronics and Design (ISLPED) 2025}}
\textit{School of Electrical and Computer Engineering, Cornell University}\\
\{cc2793, hm632, jm2787, mohamed, js3528\}@cornell.edu}

\maketitle

\begin{abstract}
Edge inference for large language models (LLM) offers secure, low-latency, and cost-effective inference solutions. We emphasize that an edge accelerator should achieve high area efficiency and minimize external memory access (EMA) during the memory-bound decode stage, while maintaining high energy efficiency during the compute-intensive prefill stage. This paper proposes an edge LLM inference accelerator featuring a hybrid systolic array (HSA) architecture that optimizes inference efficiency in both stages. To further reduce EMA, we adopt MXINT4 weight quantization and propose an optimized dataflow tailored for HSA, ensuring negligible dequantization overhead and achieving 100\% hardware utilization with minimal accuracy loss under edge DRAM bandwidth constraints. For non-linear operations, we incorporate optimized root mean square normalization (RMSNorm) and rotary position embedding (RoPE) units, reducing their latency, area, and memory access overhead while enabling end-to-end inference on our accelerator. Our solution achieves 247/117 (token/s/mm\textsuperscript{2}) while running a 1.3B LLM on long-input/long-output scenarios, providing $>$2.45$\times$/13.5$\times$ improvement over existing approaches, while maintaining superior energy efficiency in token generation.
\end{abstract}


\section{Introduction}
\label{sec:intro}
The demand for LLM inference continues to grow as a wide range of new applications are being developed. Compared to data centers, edge AI has emerged as a popular alternative, offering secure, low-power, and cost-effective inference capabilities. However, the high memory and compute demands of LLMs pose significant challenges for edge deployment.

First, edge devices have limited memory capacity, which restricts the size of deployable LLMs. Second, power consumption is a critical reliability concern, unlike in data centers where it is primarily an optimization problem. Furthermore, data centers typically benefit from batching multiple input queries to address memory bottlenecks during the decoding stage, which improves inference efficiency. In contrast, edge scenarios often involve user queries arriving one at a time, which limits the accelerator's performance due to the narrow memory bandwidth available on edge devices.

\begin{figure}[htbp]
    \centering
    \vspace{-5pt}
    \begin{subfigure}[t]{0.95\linewidth}
        \includegraphics[width=\linewidth]{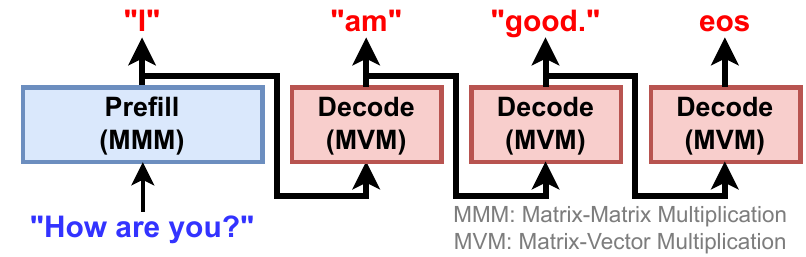}
        \vspace{-20pt}
        \caption{}
        \label{}
    \end{subfigure}
    \begin{subfigure}[t]{0.95\linewidth}
        \includegraphics[width=\linewidth]{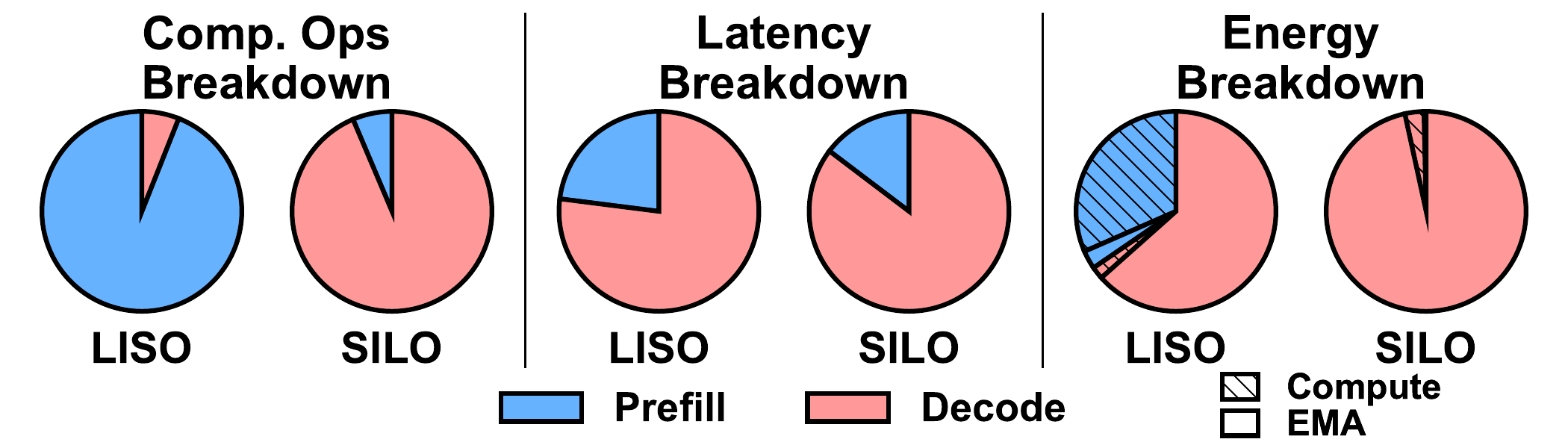}
        \vspace{-20pt}
        \caption{}
        \label{}
    \end{subfigure}
    \vspace{-10pt}
    \caption{(a) Illustration of the autoregressive token generation. (b) Breakdown of computational operations, latency, and energy for RetNet 1.3B model under LISO/SILO scenarios.} 
    \label{fig:autoregression}
\end{figure}

Fig.~\ref{fig:autoregression}(a) illustrates the autoregressive inference paradigm, where the prefill stage involves Matrix-Matrix-Multiplication (MMM), while decode comprises mainly Matrix-Vector-Multiplication (MVM) workloads. Overall, the edge inference workloads consist of two typical scenarios
of \textbf{1) Long Input Short Output (LISO)} (e.g.,~summarizing emails and extracting insights from long news articles) and \textbf{2) Short Input Long Output (SILO)} (e.g., text generation such as email templates).



For the LISO scenario, we assume input/output tokens are 750/50, while for the SILO scenario, they are 50/750. These numbers are based on the average length of an email or news article, which align with the average token statistics found in the news summarization dataset~\cite{nallapati2016abstractive}. Although in practice many input and output token lengths vary, these two settings illustrate key computational differences between prefill and decode stages under long input/output scenarios, and how edge accelerators should be optimized accordingly.

We profile the computational operations, latency, and power consumption for both LISO and SILO scenarios. We run RetNet~\cite{sun2023retentive}~1.3B LLM model on Nvidia Jetson Orin Nano~\cite{jetsonnano} as reference edge inference accelerator. 
As shown in Fig.~\ref{fig:autoregression}(b), The commercial mobile GPU achieves peak performance of 40 TOPS, but the low memory bandwidth (68 GB/s with LPDDR5) limits the speed of decoding~(only 1.7\% of peak performance), which further reveals the following observations:



\begin{enumerate}
    \item The decode stage dominates overall latency, contributing to over 80\% of the runtime in both scenarios, even when input length is significantly larger than output.
    \item EMA dominates energy consumption. Nonetheless, the computation energy consumption in the prefill stage cannot be neglected in LISO scenarios.
\end{enumerate}

These observations are consistent with results obtained using other LLMs, such as Llama~2~\cite{touvron2023llama}. Notably, edge accelerators operate under memory-bound conditions for most of the runtime, resulting in low utilization (e.g.,~$\sim$1.7\% in Jetson Orin Nano's case). Additionally, despite the low latency of the prefill stage, its energy consumption cannot be neglected in scenarios involving long inputs. Here we conclude that edge accelerators should achieve high energy efficiency during the prefill stage, while reducing EMA and maintaining area efficiency during the decode stage.

\begin{figure}[t]
    \centering
    \includegraphics[width=0.95\linewidth]{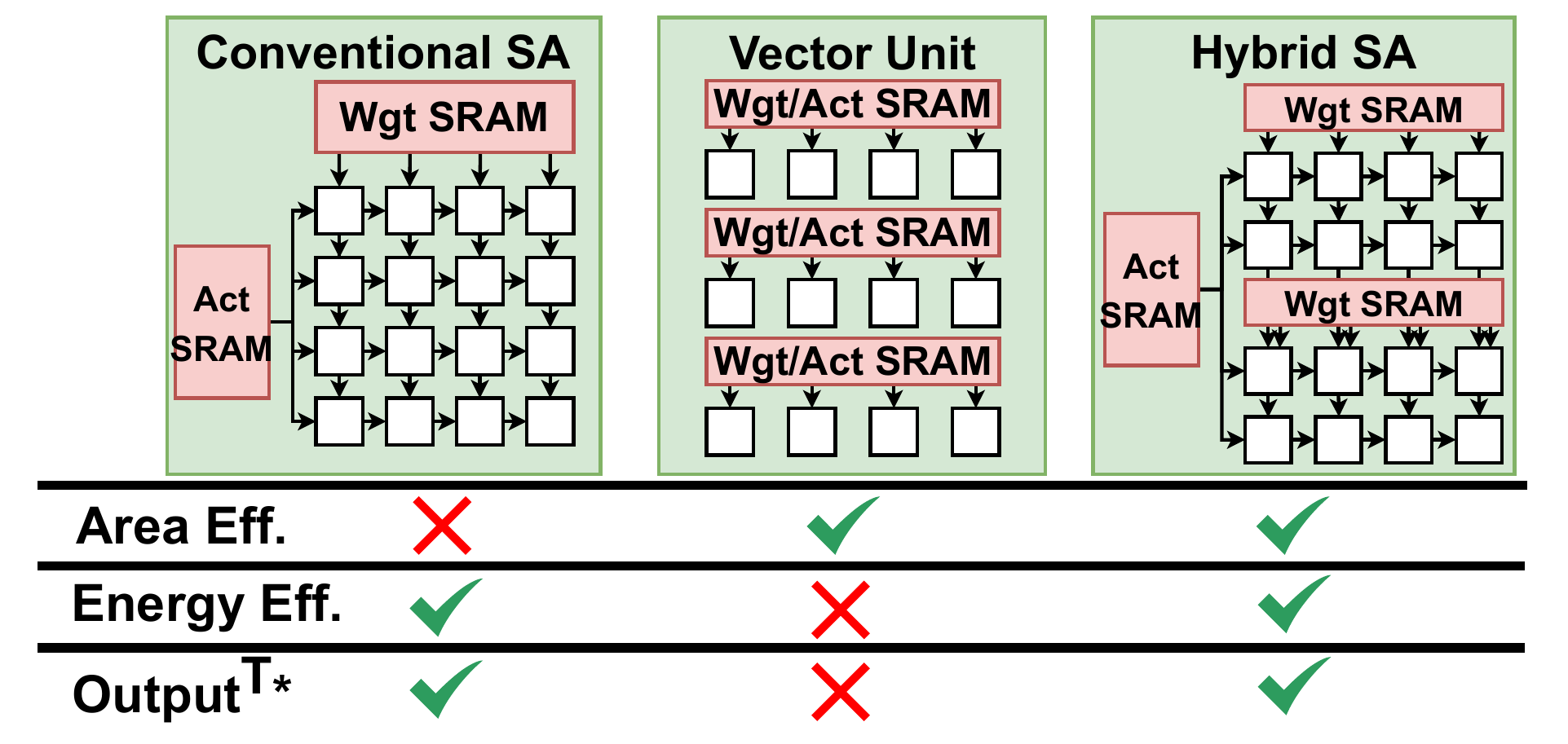}
    \caption{Comparison of three MAC architectures, highlighting the \textbf{efficiency balance challenge}. The proposed hybrid SA combines the efficiency of both SA and vector unit architectures. *Output\textsuperscript{T} indicates support for transpose output.}
    \label{fig:arch_comparison}
\end{figure}

Recent works~\cite{10.1145/3649329.3655936, keller202395, qin2023fact, qin2024mecla, islamoglu2023ita} report high peak performance and energy efficiency, but 
exhibit inefficiency when running end-to-end workloads on edge devices due to low data reuse and low utilization.
Specifically, \cite{10.1145/3649329.3655936, islamoglu2023ita, keller202395} use vector unit architecture for MAC computation, which offers flexibility to support both MMM and MVM workloads, but it fails to reuse data during prefill and leads to low energy-efficiency. In contrast, \cite{qin2023fact, qin2024mecla} employ conventional 2D systolic array (SA) architecture, which fails to support MVM workload during decode since the batch size is only one in the edge scenario, resulting in low utilization. In summary, prior architectures fail to balance energy efficiency during prefill and area efficiency during decode, as illustrated in Fig. \ref{fig:arch_comparison}.

Moreover, most of the aforementioned papers adopt INT8 as the weight format, which suffers from severe memory access bottlenecks. While \cite{keller202395} employs INT4 per-vector quantization, it fails to preserve accuracy in modern LLMs. Alternatively, \cite{qin2024mecla} introduces weight partitioning to alleviate memory bottlenecks but incurs large power and area overhead.

In this work, we present an area- and energy-efficient accelerator for edge inference, which is implemented in TSMC 28nm CMOS. We deploy RetNet 1.3B~\cite{sun2023retentive} as the target LLM to reduce memory access and eliminate the latency overhead of softmax operations. Our key contributions to address the shortcomings of prior works are as follows:
\begin{itemize}[noitemsep]
    \item \textbf{Hybrid Systolic Architecture (HSA)}: We introduce a hybrid systolic architecture that combines the strengths of conventional SA and vector units, supporting power-efficient prefilling while maintaining high hardware utilization for decoding, achieving high area efficiency.
    \item \textbf{Low-access High-utilization Dataflow}: During decode, we adopt MXINT4~\cite{rouhani2023microscaling} weight quantization scheme to reduce memory access. We propose a 100\% utilization rate dataflow mapped to HSA under 500MHz by using multiple PEs for dequantization.
    \item \textbf{Overhead-optimized RMSNorm and RoPE Units}: We introduce a layer-fused RMSNorm to eliminate latency overhead (5–10\% of total) and remove a 32kB buffer. For the RoPE unit, we reuse embedding multipliers to compute sin/cos values online, reducing DRAM access.
\end{itemize} 

Overall, our accelerator achieves 247/117 token/s/mm\textsuperscript{2} in LISO/SILO scenarios, which represents over 2.45$\times$/13.5$\times$ improvement compared to existing accelerator solutions while maintaining competitive energy efficiency. 

\begin{figure}[t!]
    \centering
    \includegraphics[width=1.05\linewidth]{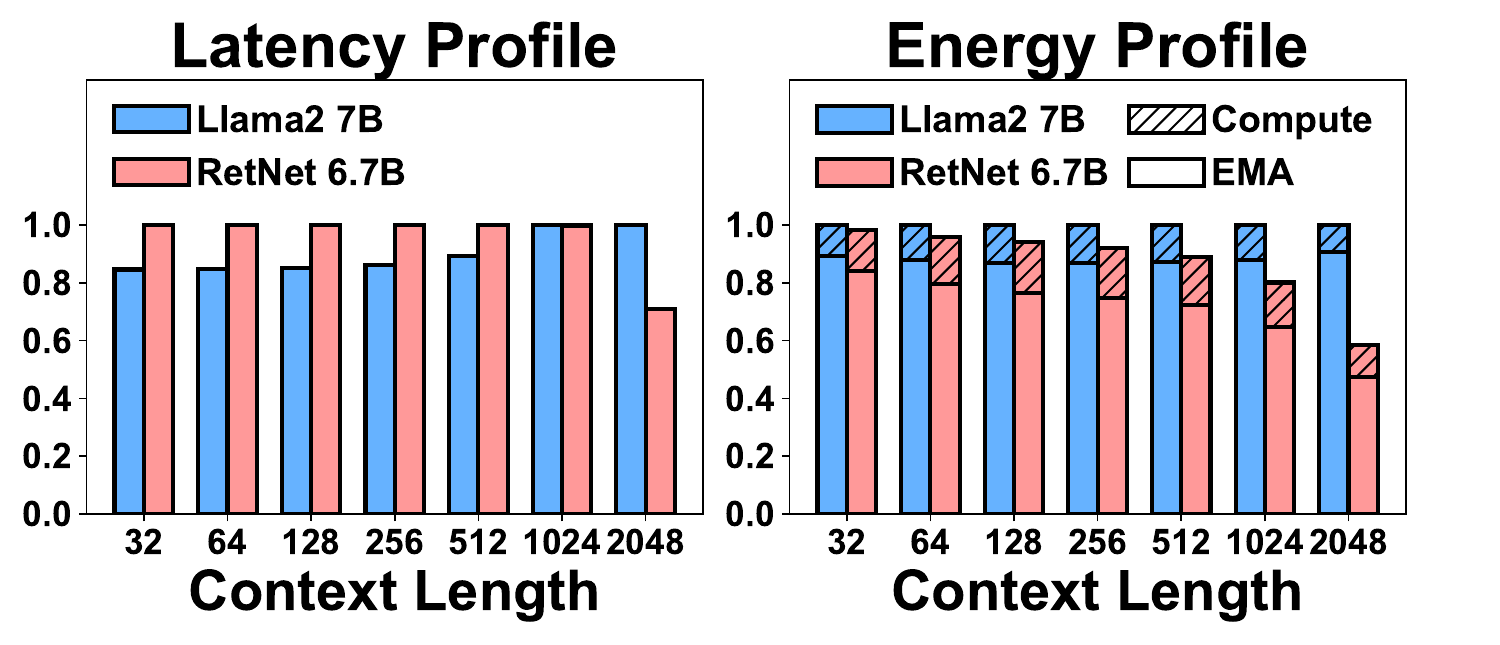}
    \vspace{-15pt}
    \caption{Normalized latency/energy comparison of Llama-2 7B and RetNet 6.7B. Profiles made with Jetson Orin Nano as a reference edge accelerator. (Energy assumption: MAC=0.5pJ/Byte \cite{jetsonnano}, DRAM access=32pJ/Byte \cite{chatterjee2017architecting,o2017fine}.)}
    \label{fig:model_comparison}
\end{figure}

\section{Background}

Retentive Network (RetNet) \cite{sun2023retentive} is a recently developed, attention-free LLM. Compared to transformer-based models, RetNet replaces the softmax operation with a decaying causal mask $D$, transforming the attention block into a retention block. This modification transforms RetNet into a Structured State Space Duality model \cite{10.5555/3692070.3692469}, providing computationally efficiency during training (similar to transformers) and enabling linear memory complexity during decoding.

\begin{figure*}[t!]
    \centering
    \includegraphics[width=\linewidth]{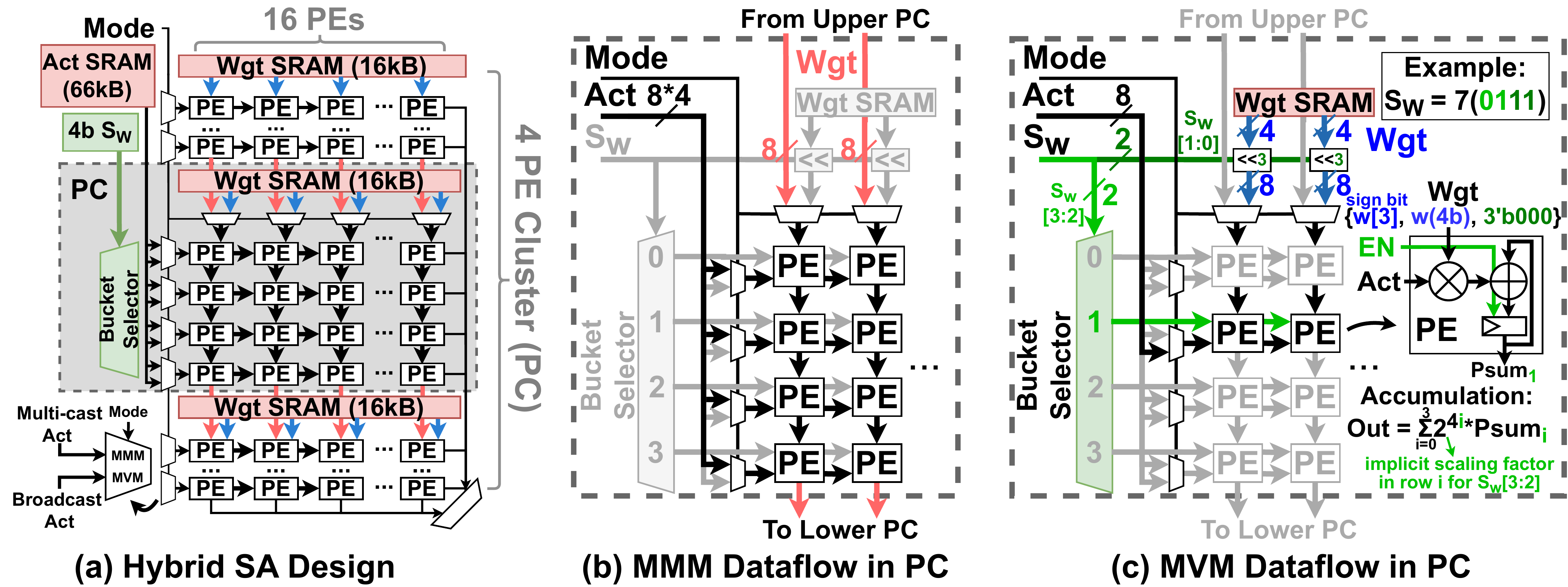}
    \caption{(a) \textbf{Left:} Proposed hybrid SA architecture. There are in total of 4 PE clusters arranged vertically with a unified activation SRAM. (b) \textbf{Middle:} MMM dataflow illustration, both Act and Wgt use INT8 format. (c) \textbf{Right:} MVM dataflow illustration, where each PC works independently in MVM since there is no weight reuse. MXINT4 format is used for Wgt.}
    \vspace{-10pt}
    \label{fig:SA}
\end{figure*}
RetNet achieves a desirable memory complexity of O(1) during decoding, making it particularly
well-suited for edge scenarios where memory bandwidth and capacity are constrained, and users can tolerate accuracy trade-offs. The energy profile in Fig.~\ref{fig:model_comparison} highlights RetNet's low memory footprint. In this profiling, we select RetNet 6.7B to match the model size of Llama~2 7B for a fair comparison.

While we chose RetNet as the target model in this work, we emphasize that all modern LLMs, including commonly used models like Llama \cite{dubey2024llama, touvron2023llama} or SLMs \cite{abdin2024phi}, employ auto-regressive token generation. Thus, the MMM (prefill) and MVM (decode) workloads remain consistent across models, all of which face the same challenges discussed in Sec.~\ref{sec:intro}.


\section{Quantization Algorithm}\label{sec:quantization}
In this work, we quantize the target RetNet~\cite{sun2023retentive} and Llama~\cite{touvron2023llama, dubey2024llama} models with 8-bit activation. The model weights are quantized to 4-bit with MXINT4 representation to reduce EMA by half during decode. 
We first adopt SmoothQuant~\cite{xiao2023smoothquant} and compress the activation precision down to INT8. 
To save the bandwidth between the off-chip memory and processing element, we follow the trend of recent state-of-the-art~(SoTA) quantization algorithms~\cite{lin2024qserve, frantar2022gptq} by quantizing the model weights into 4-bit in a group-wise fashion. 
Recent SoTA methods~\cite{lin2024qserve, frantar2022gptq, zhao2024atom} employ a floating-point 16-bit~(FP16) scaling factor as the ``dequantization scaling factor'' between the stored weights~(e.g., 4-bit) and the actual weights for computation~(e.g., 8-bit). However, performing the high-precision FP16 dequantization on the fly is expensive due to the fine-grained~(group-wise) mixed precision operation. 

In this work, given the pre-trained weight $W_{\text{FP16}}$, we adopted the scaling format from micro-scaling~(MX)~\cite{rouhani2023microscaling} by characterizing the group-wise~($W_g$) difference via a shifting-based scaling~($S_g$). Given the output channel dimension $C$ and group size $g$~($g < C$), the group-wise shifting factors are:

\vspace{-10pt}
\begin{equation}
    S_g = \text{floor}\big(\log_2(\text{max}|W_g|)\big)
\end{equation}
\vspace{-10pt}

During post-training quantization~(PTQ), the weights $W_{\text{FP16}}$ are quantized down to 4-bit after applying the shifting~(MXINT4). 
The amount of shifting is constrained between [-9, +5] to prevent overflow. On top of that, the quantization scaling factor $S_w$ remains tensor-wise, which can be fused together with the group-wise shifter~($S_g$). During prefilling, we employ INT8 weights and continue the subsequent decoding with MXINT4 format only. 
Unlike prior research works\cite{lin2024qserve, frantar2022gptq, zhao2024atom} where the shifting factors are separately stored as group-wise FP16 scaling factors, the shifting-based scaling is \textbf{naturally more compatible with hardware} compared to FP16 scaling~\cite{lin2024qserve, zhao2024atom}, while preserving minimal performance drop, as presented in Section~\ref{sec:evaluation}. Additionally, we further reduce shifting overhead by leveraging multiple PEs, taking advantage of the inherently low hardware utilization during the decode stage, as shown in  Section~\ref{sec:mvm_dataflow}.

\section{Hardware Accelerator Design}

Our accelerator consists of two main components: the Hybrid Systolic Array (HSA) and Post-Processing Unit (PPU). The HSA enables low-latency and energy-efficient computation, while PPU supports various post-processing functions to achieve end-to-end inference.
\subsection{Hybrid Systolic Array (HSA)}
The hybrid systolic array (HSA), shown in Fig.~\ref{fig:SA}(a), is designed to efficiently handle both MMM and MVM workloads. It comprises four PE clusters (PC), each with one weight SRAM, one bucket selector, 16 4-bit shifters, and 64 PEs arranged in four rows, totaling 256 PEs. All PCs share a single activation SRAM, sized to accommodate activation with a maximum dimension of 4,096 with batch size 16. While prior designs \cite{10.1145/3649329.3655936, keller202395, qin2024mecla} incorporate more than 1,024 MAC units, we opt for 256 PEs in our accelerator due to the memory-bound nature of edge inference. As highlighted in~\cite{zhang2024llmcompass}, increasing compute capability significantly helps prefill but has minimal impact and degrades utilization ratio on decoding. 

The 256 PEs are arranged as a 16$\times$16 2-D array, enabling activation and weight reuse in MMM workloads, similar to conventional SA. 
Bucket selectors and shifters are designed to incorporate MXINT4 format with 4-bit weight scaling in MVM dataflow. Furthermore, HSA supports transpose tensor manipulation by draining output vertically or horizontally. Here, we elaborate on how MMM/MVM dataflows operate on HSA architecture.

\subsubsection{\textbf{MMM Dataflow}}
Fig.~\ref{fig:SA}(b) shows the dataflow of MMM workload. As a compute-bound workload, only the weight SRAM in the first PC will be activated. Activations and weights are loaded and computed in INT8 format. Each PE adopts output stationary dataflow and is drained together at the end of operation. The draining phase consumes 16 cycles, which is negligible compared to the latency of computing large matrix multiplication. Output will be drained vertically when transposed output is required, otherwise horizontally.

\subsubsection{\textbf{MVM Dataflow}}\label{sec:mvm_dataflow}
Fig.~\ref{fig:SA}(c) shows the dataflow of MVM workload. During decode stage, as there is no reuse for weights, each PC works independently. To reduce memory access and improve hardware utilization, activations are still loaded in 8-bit while weights are loaded in 4-bit MXINT4 format. Inspired by \cite{lo2023bucket}, we map the dequantization of MXINT4 onto the HSA. As mentioned in Section III, we design our 4-bit scaling factor $S_w$ for MXINT4 in output dimension, which allows row-wise reuse. The 4-bit weights will first be shifted by two LSBs of the scaling factor ($S_w$[1:0]), resulting in 8-bit weights, and will then be transferred vertically to PEs. Two MSBs of the scaling factor ($S_w$[3:2]) activate one row of the PEs and govern accumulation of the product of 8-bit activation and 8-bit weight by clock-gating with an enable signal. At the end, output will be drained vertically and four PEs' partial sum will be accumulated together. 
Note that in MVM dataflow, we reuse the inter-PE connections in MMM dataflow to reuse existing hardware resources. The only overhead added to support MVM dataflow is the bucket selector and the shifters that cost 0.4\% of the total area.

\subsection{Post-Processing Unit (PPU)}
Our post-processing unit (PPU) supports several non-linear functions, such as activation functions and quantization, after the MAC is conducted to enable end-to-end inference. We particularly optimized RMSNorm and RoPE units, which have larger area, latency, and energy overhead.

\begin{figure}[t!]
    \centering
    \includegraphics[width=\linewidth]{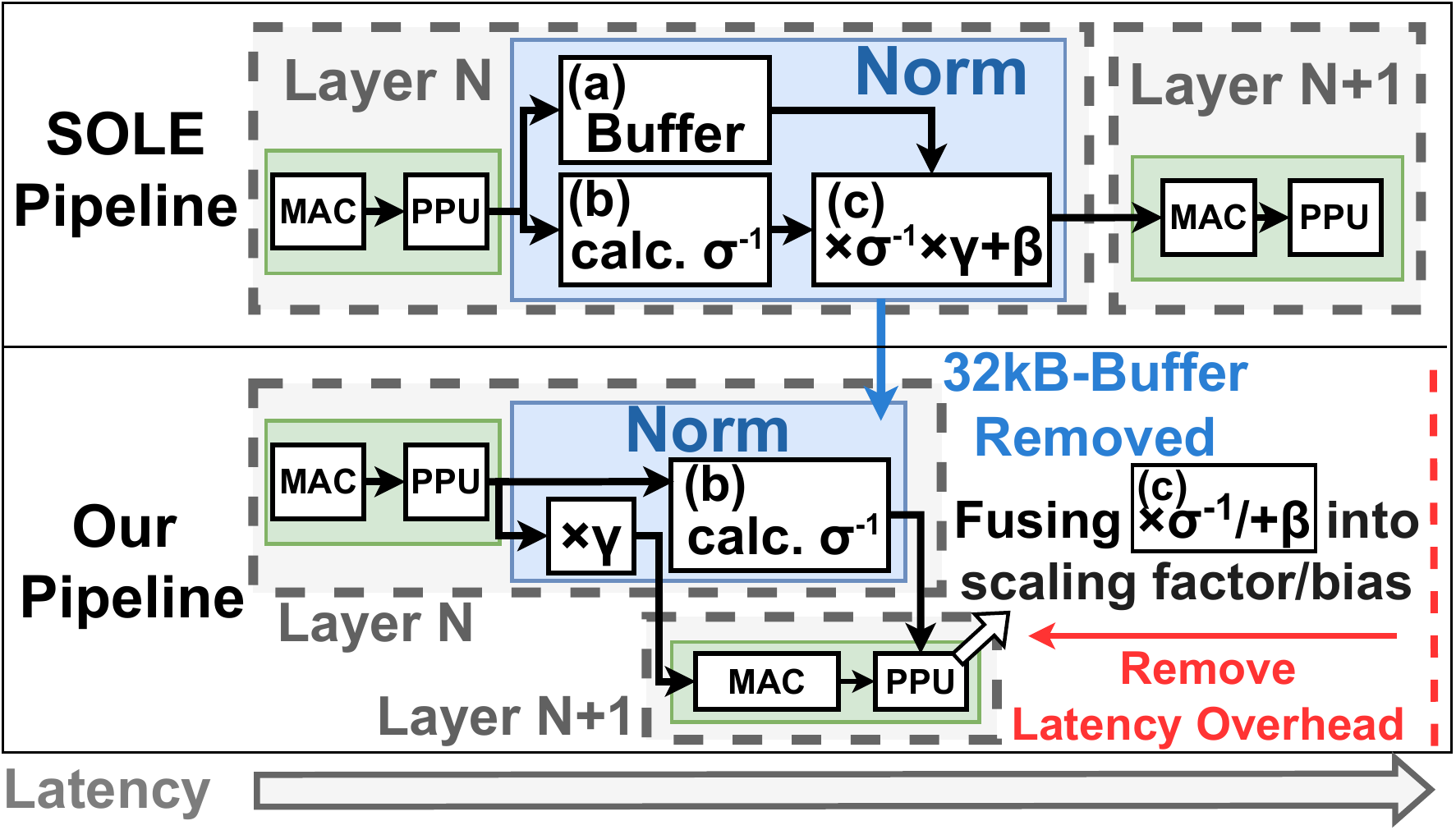}
    \vspace{-10pt}
    \caption{Comparison of normalization pipeline between SOLE~\cite{wang2023sole} and ours.}
    \label{fig:norm}
\end{figure}

\subsubsection{Root Mean Square Normalization (RMSNorm)}

LLM models adopt RMSNorm over standard normalization to eliminate the latency and area overhead caused by mean computation, which further simplifies the design of AI accelerators.

Prior works~\cite{wang2023sole} implemented a dedicated module for standard normalization which contains (a) input buffer, (b) $\sigma^{-1}$ calculation, and (c) normalization with affine transform. Since our focus is solely on RMSNorm, we propose a fusion scheme that reduces area overhead by removing the need of the input buffer, while eliminating latency overhead by parallelly calculating $\sigma^{-1}$ and the next layer.

Eq.~\eqref{eq:mac} describes a quantized matrix multiplication layer~$n$ where $Y_n $, $X_n$, $W_n$, $S_n$ denote the output activation, input activation, weight, and quantization scaling factor respectively. Eq.~\eqref{eq:rmsnorm} describes the RMSNorm calculation, where $\sigma_{Y_n}$, $\gamma$, and $\beta$ are the standard deviation and learned parameters. Note that $\gamma$ and $\beta$ are vectors of the same dimension as $Y_n$.

\vspace{-10pt}
\begin{equation}
    Y_n=(X_n\cdot W_n)\cdot S_n
    \label{eq:mac}
\end{equation}
\vspace{-10pt}
\begin{equation}
    RMSNorm(Y_n)=\frac{Y_n}{\sigma_{Y_n}}*\gamma+\beta
    \label{eq:rmsnorm}
\end{equation}
\vspace{-10pt}

Instead of performing the full RMSNorm on $Y_n$ in layer $n$, we apply only $*\gamma$ and fuse $*\sigma^{-1}_{Y_n}$ and $+\beta$ with the scaling factor $S_{n+1}$ and bias $B_{n+1}$ in layer $n+1$ as shown in Eq.~\eqref{eq:fuse_norm}. Bold terms are computed online. This fusion preserves mathematical equivalence and enables pipelining. $\beta$ is included for generality, although it is often omitted in modern LLMs.

\begin{equation}
    \begin{aligned}
        \mathbf{X_{n+1}}=&\;RMSNorm(\mathbf{Y_n})\\
        =&\;\mathbf{Y_n}*\mathbf{\sigma^{-1}_{Y_n}}*\gamma+\beta\\
        \mathbf{Y_{n+1}}=&\;(W_{n+1}*\mathbf{X_{n+1}})*S_{n+1}\\
        =&\;W_{n+1}*(\mathbf{Y_n}*\mathbf{\sigma^{-1}_{Y_n}}*\gamma+\beta)*S_{n+1}\\
        =&\;W_{n+1}*\mathbf{Y_n}*\gamma*\mathbf{\sigma^{-1}_{Y_n}}*S_{n+1}+W_{n+1}*\beta\\
         &\;*S_{n+1}\\
        =&\;W_{n+1}*\mathbf{Y_n^*}*\mathbf{S_{n+1}^*}+B_{n+1}\\ 
        \text{where }&\mathbf{Y_n^*=Y_n*\gamma}, \;\;\mathbf{S_{n+1}^*}=\mathbf{\sigma^{-1}_{Y_n}}*S_{n+1}, \\
        &B_{n+1}=W_{n+1}*\beta*S_{n+1}
    \end{aligned}
    \label{eq:fuse_norm}
\end{equation}
\vspace{-5pt}

Fig.~\ref{fig:norm} illustrates the overhead optimization compared to SOLE. We effectively remove the (a) 32kB buffer (assuming embedding dimension of 4,096, batch size of 16) in RMSNorm unit and eliminate its latency overhead (5-10\% of total latency) by pipelining its operations (b) with the subsequent layer’s MAC operations. The (b) calculation of $\sigma^{-1}$ remains the same, which involves square accumulation and square root operation.

\subsubsection{Rotary Position Embedding (RoPE)}
RoPE \cite{su2024roformer} is widely adopted to incorporates token position information in LLMs. In RoPE, for the $m^{th}$ token, its embeddings $x_n$ and $x_{n+1}$ are rotated by an angle of $m\theta_{n/2}$. The embedding computation is shown in Eq.~\eqref{eq:rope}, where $d$ represents the embedding dimension per head.

\vspace{-5pt}
\begin{equation}
    \begin{aligned}
    & RoPE(x_n, x_{n+1}) = \\
    & \begin{pmatrix}
        x_n \\
        x_{n+1}
    \end{pmatrix}
    \otimes
    \begin{pmatrix}
        \cos m\theta_{n/2} \\
        \cos m\theta_{n/2}
    \end{pmatrix}
    +
    \begin{pmatrix}
        -x_{n+1} \\
        x_n
    \end{pmatrix}
    \otimes
    \begin{pmatrix}
        \sin m\theta_{n/2} \\
        \sin m\theta_{n/2} 
    \end{pmatrix} \\
    & \qquad\qquad \theta_i=10000^{-2(i-1)/d}, i \in [1, 2, ..., d/2]
    \end{aligned}
    \label{eq:rope}
\end{equation}

Na\"ive implementations of RoPE either involve loading pre-computed sine/cosine values as weights, or compute them on the accelerator. Consider the memory-bound nature of LLM inference and the high cost of on-chip CORDIC module, our RoPE unit reuses the existing embedding multipliers and adders to compute sine and cosine values for the next token with trigonometric identities~(Eq.~\eqref{eq:trigon}). This approach reuses the hardware resources and also eliminates the need for repeated DRAM access.
\begin{figure}[t!]
    \centering
    \includegraphics[width=\linewidth]{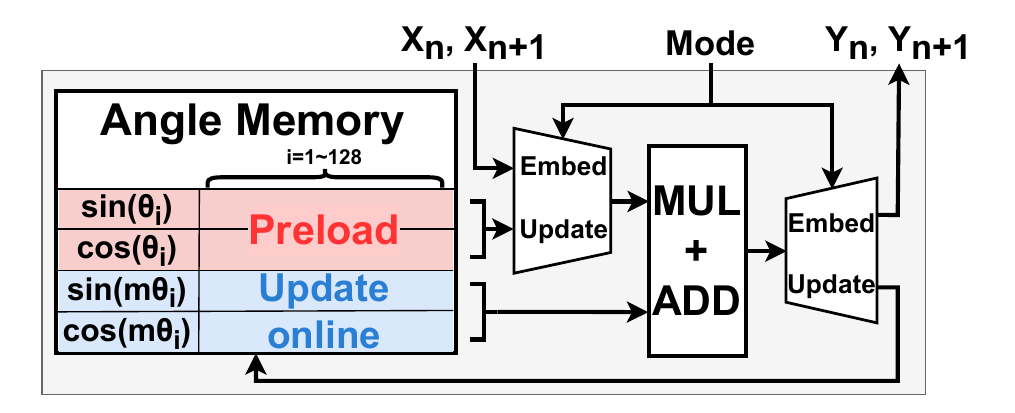}
    \vspace{-15pt}
    \caption{Proposed RoPE unit. Angle memory stores sin, cos values across head dimension $d$. RoPE has two modes of operation: \textbf{Embed} mode and \textbf{Update} mode.}
    \label{fig:rope}
\end{figure}

\begin{equation}
    \begin{aligned}
    sin((m+1)\theta_i)=sin(m\theta_i)cos(\theta_i)+sin(\theta_i)cos(m\theta_i)\\
    cos((m+1)\theta_i)=cos(m\theta_i)cos(\theta_i)-sin(m\theta_i)sin(\theta_i)
    \end{aligned}
    \label{eq:trigon}
\end{equation}

Fig.~\ref{fig:rope} illustrates the details of the RoPE unit. While performing rotary position embedding for token $m$, the unit operates in \textbf{Embed} mode, where token embeddings $x_n, x_{n+1}$ are embedded with $sin(m\theta_i)$ and $cos(m\theta_i)$. 
Subsequently, the unit switches to \textbf{Update} mode to calculate $sin((m+1)\theta_i)$ and $cos((m+1)\theta_i)$ using trigonometric identities, preparing for the the next token $m+1$. 
The RoPE unit acquires $\sim$4\% of the total area of our accelerator.

\section{Experimental Results}\label{sec:evaluation}
\begin{figure}[t!]
    \centering
    \includegraphics[width=\linewidth]{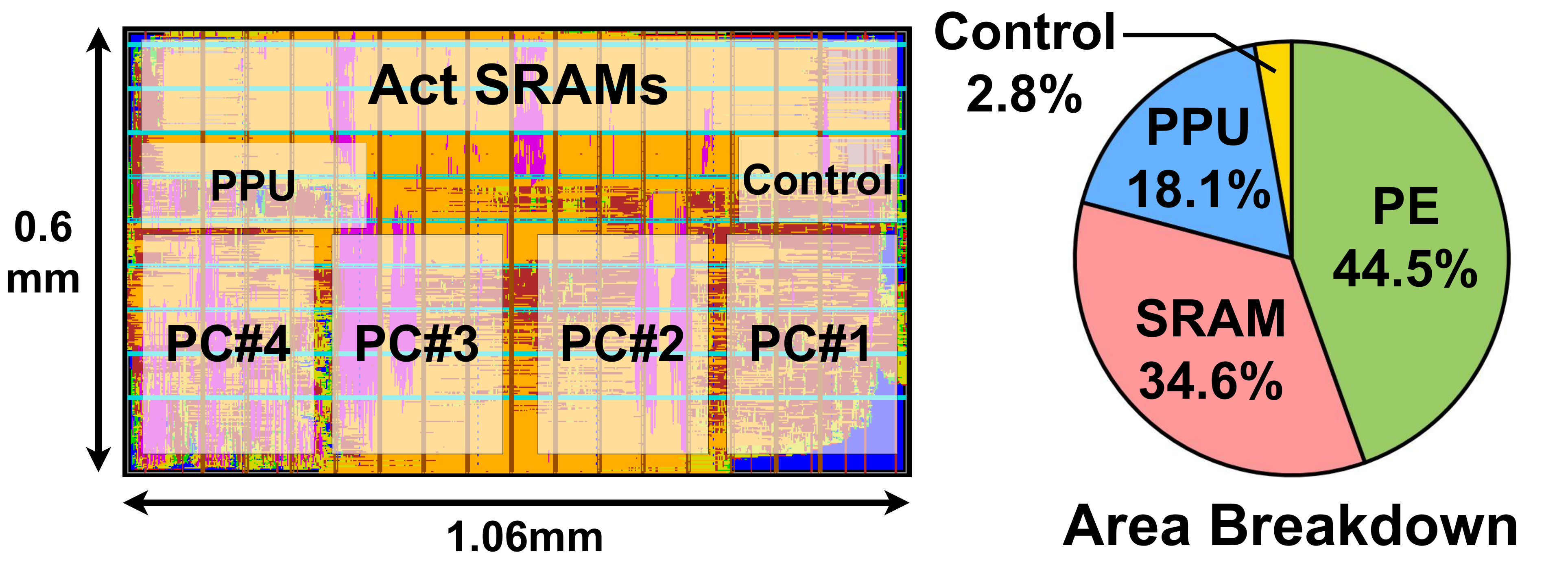}
    \vspace{-15pt}
    \caption{28nm accelerator layout and area breakdown.}
    \label{fig:layout}
\end{figure}

Our accelerator is prototyped in TSMC 28nm CMOS and is evaluated with post-layout simulation results. The design is implemented in SystemC code following the high level synthesis (HLS) design flow \cite{khailany2018modular} with Siemens Catapult. 
For gate-level design, Synopsys Design Compiler is used to synthesize RTL code from Catapult, where clock gating is added to the design. Cadence Innovus is used as the place-and-route tool. To obtain the power and latency results with post-layout simulation, we use Synopsys VCS and Primetime. Fig.~\ref{fig:layout} shows the 28nm accelerator layout and area breakdown.

\subsection{Architecture Comparison 
}

To highlight HSA's advantages over conventional SA and vector unit, we compare their end-to-end energy and latency while running RetNet 1.3B under LISO and SILO scenarios. HSA is evaluated using post-layout simulation, while \cite{keller202395} scaled to 28nm as the vector unit baseline. Conventional SA performance is estimated based on HSA, as it introduces minimal overhead. Note that as both prefill and decode stages are evaluated, ``token'' refers to total tokens, including prompt and output, in token-related metrics like tokens/s. 

\begin{figure}[t!]
    \centering
    \includegraphics[width=\linewidth]{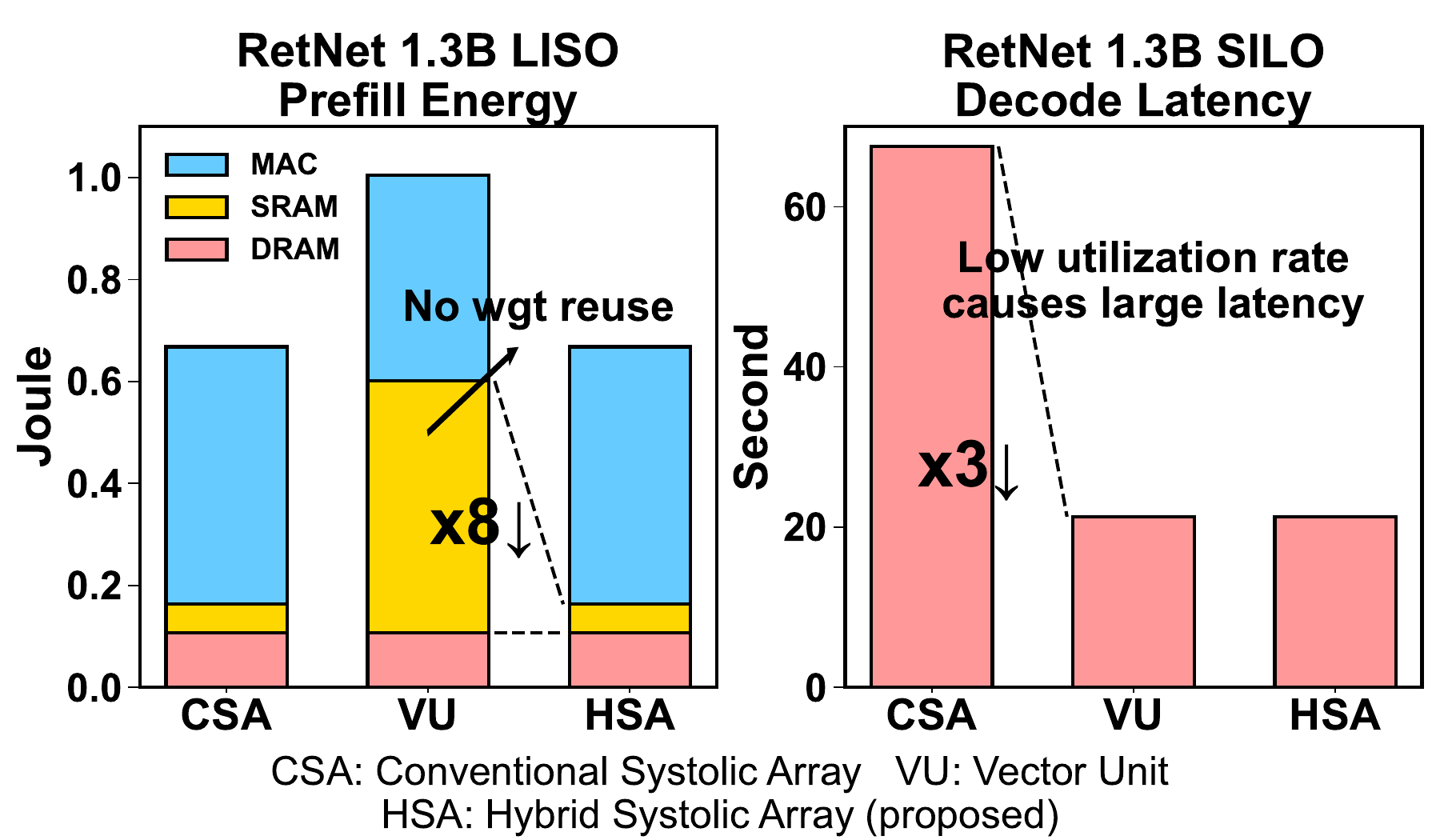}
    \vspace{-15pt}
    \caption{Energy and latency in prefill and decoding.}
    \label{fig:arch_eval-stage}
\end{figure}

\begin{table}[ht]
\renewcommand{\arraystretch}{1.04}
\centering
\caption{Performance comparison of different architectures. }
\begin{tabular}{cc|ccc}
\hline
\multicolumn{2}{c|}{\textbf{Architecture}}                                                   & \textbf{Conv.~SA\textsuperscript{a}} & \textbf{Vector Unit\textsuperscript{b}}                               & \textbf{HSA\textsuperscript{c}}                                                   \\ \hline
\multicolumn{1}{c|}{\multirow{2}{*}{Tokens/s}}                     & LISO\textsuperscript{d}                     & 90.2         & 138.3                                              & \textbf{138.3}                                                 \\
\multicolumn{1}{c|}{}                                             & SILO\textsuperscript{e}                     & 11.8         & 37.6                                               & \textbf{37.6}                                                  \\ \hline
\multicolumn{1}{c|}{\multirow{2}{*}{Tokens/J}}                     & LISO\textsuperscript{d}                     & 1060.7       & 719.1                                              & \textbf{1060.7}                                                \\
\multicolumn{1}{c|}{}                                             & SILO\textsuperscript{e}                     & 21.83        & 21.6                                               & \textbf{21.83}                                                 \\ \hline
\multicolumn{2}{c|}{\begin{tabular}[c]{@{}c@{}}Low-bit \\ quantization support\end{tabular}} & None         & \begin{tabular}[c]{@{}c@{}}VSQ\\ W4A4\end{tabular} & \textbf{\begin{tabular}[c]{@{}c@{}}MXINT4\\ W4A8\end{tabular}} \\ \hline 
\multicolumn{5}{l}{Hardware Settings: (1) 256 PEs   (2) 500MHz   (3) Same tiling strategy}\\
\multicolumn{5}{l}{LLM Settings: End-to-end RetNet 1.3B}\\
\hline
\multicolumn{5}{l}{\textsuperscript{a}Estimated based on HSA by considering the dataflow differences} \\
\multicolumn{5}{l}{\textsuperscript{b}The result of~\cite{keller202395} is scaled to 28nm CMOS as vector unit baseline} \\
\multicolumn{5}{l}{\textsuperscript{c}Post-layout simulation results} \\
\multicolumn{5}{l}{\textsuperscript{d,e}Prompt/output tokens is 750/50 for LISO, 50/750 for SILO} 
\end{tabular}
\label{table:arch-table}
\end{table}

\begin{table*}[t!]
\renewcommand{\arraystretch}{1.0}
\centering
\caption{{Comparison of proposed accelerator and SoTA works.}}

\begin{tabular}{ccccccc}
\hline
\multicolumn{1}{|c|}{}                                                                           & \multicolumn{1}{c|}{\begin{tabular}[c]{@{}c@{}}Keller et al.\textsuperscript{a} \\ JSSC'23\cite{keller202395}\end{tabular}}           & \multicolumn{1}{c|}{\begin{tabular}[c]{@{}c@{}}MECLA\textsuperscript{b}\\ ISCA'24\cite{qin2024mecla}\end{tabular}} & \multicolumn{1}{c|}{\begin{tabular}[c]{@{}c@{}}FACT\textsuperscript{c}\\ ISCA'23\cite{qin2023fact}\end{tabular}} & \multicolumn{1}{c|}{\begin{tabular}[c]{@{}c@{}}Kim et al.\textsuperscript{d}\\ ISSCC'24\cite{kim202420}\end{tabular}} & \multicolumn{1}{c|}{\begin{tabular}[c]{@{}c@{}}Moon et al.\textsuperscript{d}\\ ISSCC'23\cite{moon2023127}\end{tabular}} & \multicolumn{1}{c|}{\textbf{This work}}                                     \\ \hline
\multicolumn{1}{|c|}{Process Tech (nm)}                                                          & \multicolumn{1}{c|}{5 (28)}                                                                        & \multicolumn{1}{c|}{28}                                                      & \multicolumn{1}{c|}{28}                                                         & \multicolumn{1}{c|}{28}                                                                & \multicolumn{1}{c|}{28}                                                                 & \multicolumn{1}{c|}{28}                                                     \\ \hline
\multicolumn{1}{|c|}{Dataflow}                                                                   & \multicolumn{1}{c|}{MVM}                                                                           & \multicolumn{1}{c|}{MMM}                                                     & \multicolumn{1}{c|}{MMM}                                                        & \multicolumn{1}{c|}{MMM}                                                               & \multicolumn{1}{c|}{MVM}                                                                & \multicolumn{1}{c|}{\textbf{Hybrid}}                                        \\ \hline
\multicolumn{1}{|c|}{Area (mm\textsuperscript{2})}                                                                 & \multicolumn{1}{c|}{0.153 (4.8)}                                                                   & \multicolumn{1}{c|}{22.02}                                                   & \multicolumn{1}{c|}{6.03}                                                       & \multicolumn{1}{c|}{20.25}                                                             & \multicolumn{1}{c|}{7.29}                                                               & \multicolumn{1}{c|}{\textbf{0.636}}                                         \\ \hline
\multicolumn{1}{|c|}{Frequency (MHz)}                                                            & \multicolumn{1}{c|}{152-1760}                                                                      & \multicolumn{1}{c|}{1000}                                                    & \multicolumn{1}{c|}{500}                                                        & \multicolumn{1}{c|}{50-200}                                                            & \multicolumn{1}{c|}{20-400}                                                             & \multicolumn{1}{c|}{500}                                                    \\ \hline
\multicolumn{1}{|c|}{Power (W)}                                                                  & \multicolumn{1}{c|}{Not Reported}                                                                  & \multicolumn{1}{c|}{2.87}                                                    & \multicolumn{1}{c|}{0.337}                                                      & \multicolumn{1}{c|}{Not Reported}                                                      & \multicolumn{1}{c|}{0.002-0.237}                                                        & \multicolumn{1}{c|}{\textbf{0.108}}                                         \\ \hline
\multicolumn{1}{|c|}{Post Proccesing Unit}                                                       & \multicolumn{1}{c|}{YES}                                                                           & \multicolumn{1}{c|}{YES}                                                     & \multicolumn{1}{c|}{YES}                                                        & \multicolumn{1}{c|}{NO}                                                                & \multicolumn{1}{c|}{NO}                                                                 & \multicolumn{1}{c|}{\textbf{YES}}                                           \\ \hline
\multicolumn{1}{|c|}{Data Format}                                                                & \multicolumn{1}{c|}{INT4/INT8}                                                                     & \multicolumn{1}{c|}{INT8}                                                    & \multicolumn{1}{c|}{INT8}                                                       & \multicolumn{1}{c|}{INT8}                                                              & \multicolumn{1}{c|}{AQ 1-8b}                                                            & \multicolumn{1}{c|}{\textbf{MXINT4/INT8}}                                   \\ \hline
\multicolumn{1}{|c|}{Peak Performance (TOPS)}                                                    & \multicolumn{1}{c|}{3.6/1.8}                                                                       & \multicolumn{1}{c|}{14}                                                      & \multicolumn{1}{c|}{1.02}                                                       & \multicolumn{1}{c|}{3.41}                                                              & \multicolumn{1}{c|}{0.52}                                                               & \multicolumn{1}{c|}{0.256}                                                  \\ \hline
\multicolumn{1}{|c|}{\begin{tabular}[c]{@{}c@{}}Peak Energy Efficiency \\ (TOPS/W)\end{tabular}} & \multicolumn{1}{c|}{\begin{tabular}[c]{@{}c@{}}91.1(2.9)/39.1(1.25) \\ @0.47V,152MHz\end{tabular}} & \multicolumn{1}{c|}{7.08}                                                    & \multicolumn{1}{c|}{4.388}                                                      & \multicolumn{1}{c|}{\begin{tabular}[c]{@{}c@{}}22.9 \\ @0.5V,50MHz\end{tabular}}       & \multicolumn{1}{c|}{\begin{tabular}[c]{@{}c@{}}8.94 \\ @0.46V,20MHz\end{tabular}}       & \multicolumn{1}{c|}{\begin{tabular}[c]{@{}c@{}}2.37 \\ @0.81V\end{tabular}} \\ \hline
\multicolumn{1}{|c|}{LISO\textsuperscript{e} Area Effi. (token/s/mm\textsuperscript{2})\textsuperscript{f}}                                        & \multicolumn{1}{c|}{395.18 (82.36)\textsuperscript{g}}                                                           & \multicolumn{1}{c|}{100.82}                                                  & \multicolumn{1}{c|}{50.83}                                                     & \multicolumn{1}{c|}{22.84}                                                             & \multicolumn{1}{c|}{30.83}                                                              & \multicolumn{1}{c|}{\textbf{247.38}}                                        \\ \hline
\multicolumn{1}{|c|}{SILO\textsuperscript{e} Area Effi. (token/s/mm\textsuperscript{2})\textsuperscript{f}}                                        & \multicolumn{1}{c|}{38.00(7.92)\textsuperscript{g}}                                                              & \multicolumn{1}{c|}{8.63}                                                   & \multicolumn{1}{c|}{6.29}                                                      & \multicolumn{1}{c|}{1.88}                                                              & \multicolumn{1}{c|}{5.19}                                                              & \multicolumn{1}{c|}{\textbf{116.55}}                                        \\ \hline
\multicolumn{1}{|c|}{{Prefill Energy}  w/ EMA (mJ/token)\textsuperscript{f}}                                  & \multicolumn{1}{c|}{0.246\textsuperscript{g}}                                                                     & \multicolumn{1}{c|}{0.449}                                                   & \multicolumn{1}{c|}{0.685}                                                      & \multicolumn{1}{c|}{0.187}                                                             & \multicolumn{1}{c|}{0.148}                                                              & \multicolumn{1}{c|}{0.773}                                                  \\ \hline
\multicolumn{1}{|c|}{{Decode Energy}  w/ EMA (mJ/token)\textsuperscript{f}}                                   & \multicolumn{1}{c|}{48.59\textsuperscript{g}}                                                                     & \multicolumn{1}{c|}{25.10}                                                   & \multicolumn{1}{c|}{54.54}                                                      & \multicolumn{1}{c|}{47.96}                                                             & \multicolumn{1}{c|}{45.15}                                                              & \multicolumn{1}{c|}{\textbf{24.06}}                                         \\ \hline
\multicolumn{7}{l}{\textsuperscript{a}Scaled results show in parentheses for fair comparison across different technology node.}                                                                                                                                                                                                                                                                                                                                                                                                                                                                                                                        \\
\multicolumn{7}{l}{\textsuperscript{b}TOPS reported is optimized by fine-tuning weight compression algorithm with accuracy loss, using SSMP standard setting}                                                                                                                                                                                                                                                                                                                                                                                                                                                                                                                       \\
\multicolumn{7}{l}{\textsuperscript{c}TOPS reported is optimized by exploiting sparsity with accuracy loss \quad \quad \quad \textsuperscript{d}Results without applying sparsity}                                                                                                                                                                                                                                                                                                                                                                                                                                                                                                      \\
\multicolumn{7}{l}{\textsuperscript{e}Running on RetNet 1.3B with prompt/output tokens of 750/50 for LISO and 50/750 for SILO}                                                                                                                                                                                                                                                                                                                                                                                                                                                                                                                         \\
\multicolumn{7}{l}{\textsuperscript{f}Using DDR5 51.2GB/s bandwidth to simulate accelerator's behavior during edge decoding. DRAM access energy consumption 32pJ/Byte}                                                                                                                                                                                                                                                                                                                                                                                                                                                                                                                        \\
\multicolumn{7}{l}{{\textsuperscript{g}INT4-VSQ datapath cannot maintain acceptable model accuracy based on our experiment, results shown are using INT8 datapath}}                                                                                                                                                                                                                                                                                                                                                                                                                                                                                     
\end{tabular}

\vspace{-10pt}
\label{table:eval-table}
\end{table*}

Fig.~\ref{fig:arch_eval-stage} presents energy and latency results for the prefill and decode stages. We report only prefill energy under LISO scenario and decode latency under SILO scenario for two reasons: 1) Decode energy remains consistent due to excessive DRAM access, while LISO's large prefill computation highlights architectural differences. 2) During prefill, accelerators are compute-bound, making latency differences negligible for designs with same PE counts. In contrast, The SILO scenario better captures variations in decode latency. Our evaluation shows that vector unit consume 36\% more energy than the other two architectures due to excessive SRAM access, as there is no weight reuse (activations can be reused by register buffers). In terms of latency, conventional SA suffers from low utilization rate, leading to poor decoding latency.

Table~\ref{table:arch-table} presents the end-to-end performance comparison of the three architectures. HSA reducing power consumption by reusing activations and weights, while maintaining high utilization. These results demonstrate its effectiveness of addressing efficiency balance challenge mentioned in Fig. \ref{fig:arch_comparison}.

\subsection{Comparison with Prior Works}

Table~\ref{table:eval-table} shows the comparison with other works, including Keller et al. \cite{keller202395}, FACT \cite{qin2023fact}, MECLA \cite{qin2024mecla}, and other state-of-the-art ASIC accelerators \cite{kim202420, moon2023127}. While these SoTA works report high `peak' performance (TOPS) and energy efficiency (TOPS/W), memory bandwidth limitation in edge scenarios often constrain the accelerator from achieving such peak during inference. To reflect real-world edge conditions, we evaluate our design using end-to-end RetNet 1.3B inference with DDR5 memory bandwidth limitations (51.2GB/s).

Our accelerator achieves 2.45-10.83$\times$ and 13.5-61.99$\times$ improvements in area efficiency for token generation (token/s/mm\textsuperscript{2}) under LISO and SILO, respectively. This gain primarily stems from the high utilization of HSA and the integration of low-bit quantization to mitigate memory bottlenecks. Our accelerator maintains superior area efficiency across various token length settings, with LISO and SILO serving as representative evaluation scenarios for long context.

\begin{table}[t!]

\caption{{WikiText2~\cite{merity2016pointer} perplexity$\downarrow$ with 2,048 text length.} }
\resizebox{\linewidth}{!}{
\begin{tabular}{c|c|cccc}
\hline
Method             & Precision & Scaling           & RetNet-3B & Llama 3.2-3B & Llama 2-7B \\ \hline
Baseline           & FP16      & -                 & 16.58     & 11.66        & 5.47       \\ \hline
SmoothQ.\cite{xiao2023smoothquant}           & W8A8      & FP16                 & 17.97     & 11.79        & 5.54       \\ \hline
Atom\cite{zhao2024atom}               & W4A8      & FP16              & -         & -            & 5.91       \\ \hline
QServe\cite{lin2024qserve}             & W4A8      & FP16              & -         & -            & 5.70       \\ \hline
\multirow{2}{*}{VSQ\cite{dai2021vs}}                & W8A8      & INT8         &      -     & 11.736       & 5.44       \\ \cline{2-6} 
                   & W4A8      & INT8         &   -        & 1e35         & 107329     \\ \hline
\textbf{This work} & W4A8      & \textbf{4b Shift} & 18.22     & 13.11        & 5.81       \\ \hline
\end{tabular}
}
\label{tbl:ppl}
\end{table}

\begin{table}[t!]

\caption{{Generative reasoning accuracy$\uparrow$ of GSM8K.} }
\resizebox{\linewidth}{!}{
\begin{tabular}{c|c|ccc}
\hline
Method             & Precision & Scaling & Llama 3.1-8B-Instruct & Llama 3.2-3B-Instruct \\ \hline
Baseline           & FP16      & -                 & 85.45       & 75.22       \\ \hline
SmoothQ.\cite{xiao2023smoothquant}           & W8A8      & FP16     & 83.11        & 71.42       \\ \hline
SmoothQ.\cite{xiao2023smoothquant}           & W4A8      & FP16     & 2.54        & 10.56        \\ \hline
\textbf{This work} & W4A8      & \textbf{4b Shift}     & 82.54        & 69.22       \\ \hline
\end{tabular}
}
\label{tbl:gsm8k}
\end{table}

For energy efficiency, reducing EMA is crucial, as shown in Fig.~\ref{fig:autoregression}. While \cite{keller202395, dai2021vs} supports 4-bit per-vector quantization, our results (Table~\ref{tbl:ppl}) show that it fails to sustain performance under W4A8. \cite{qin2024mecla} proposes a weight partitioning technique that successfully reduces EMA by 80\% but introduces significant area and operating power overhead--both critical concerns for edge devices--and also requires fine-tuning on models.

In contrast, our approach reduce EMA by adopting MXINT4 weight quantization, maintaining high accuracy while being plug-and-play solution for any LLM. Our accelerator consumes 24.06 mJ/token for RetNet 1.3B token generation. 
The energy consumption reported in~\cite{kim202420} 
was evaluated on Wikitext2~\cite{merity2016pointer}, which only goes though prefill stage and generates one token, whereas real-world applications decode numerous tokens and decoding dominates overall energy cost.

Although our evaluation focuses on the RetNet model with certain token length use cases, the high utilization rate and improvements in area and energy efficiency can be retained when deploying other commonly used models like Llama and in different token length scenarios, as modern LLMs share the same core computations—MMM and MVM.

\begin{table}[t!]
\renewcommand{\arraystretch}{1.04}
\centering
\caption{Dequantization hardware overhead.}
\begin{tabular}{|c|ccc|}
\hline
Dequant. Scaling & 4-bit Shift & INT8  & FP16  \\ \hline
Normalized Area     & \textbf{1.0$\times$}           & 10.33$\times$ & 15.96$\times$ \\ \hline
Normalized Power    & \textbf{1.0$\times$}           & 7.19$\times$  & 9.60$\times$  \\ \hline
\end{tabular}
\label{tab:dequant}
\end{table}

\subsection{Accuracy with Proposed Quantization Scheme}
We evaluate the proposed hardware-friendly quantization scheme (Section~\ref{sec:quantization}) for both RetNet and transformer-based Llama LLMs. We choose the weight group size to 16~(along the output channel) to match the capacity of each PE of the accelerator. Table~\ref{tbl:ppl} and Table~\ref{tbl:gsm8k} show the comparison of perplexity results on WikiText2 dataset~\cite{merity2016pointer} and GSM8K dataset with SoTA quantization schemes~\cite{xiao2023smoothquant, lin2024qserve}.
With 4-bit weight and 8-bit activation~(W4A8), the proposed quantization method achieves performance on par with recent SoTA quantization methods with 8-bit weight and 8-bit activation (W8A8). For the reasoning-based GSM8K dataset, we evaluate our proposed method with instruction-tuned Llama model only since the RetNet is not designed for reasoning. 

\section{Conclusion}
This work presents an efficient LLM accelerator 
featuring the hybrid systolic array architecture, which addresses the efficiency balance challenge in edge devices. We further reduce EMA by adopting MXINT4 format on weights, which is mapped onto our HSA architecture with high hardware utilization. 
Evaluated on post-layout simulation results with 500 MHz operating frequency, our accelerator achieves $>$2.45$\times$/13.5$\times$ higher area efficiency across LISO/SILO scenarios for end-to-end token generation, 
while maintaining high energy efficiency in token generation.

\section*{Acknowledgment}
This work is supported in part by NSF under grant 2303626, and CoCoSys Center in SRC/DARPA JUMP 2.0 program. 

\bibliographystyle{IEEEtranS}
\bibliography{references}

\end{document}